\documentclass[twocolumn,showpacs,groupedaddress,amsmath,amssymb]{revtex4}

\usepackage{graphicx}
\usepackage{dcolumn}
\usepackage{bm}
\usepackage{amsmath,amssymb}

\begin{document}

\preprint{APS/123-QED}

\title{Ferromagnetic interaction between Cu ions in the bulk region of Cu-doped ZnO nanowires}

\author{T.~Kataoka, Y.~Yamazaki, V.~R.~Singh, and A.~Fujimori}
\affiliation{Department of Physics and Department of Complexity Science and Engineering, 
University of Tokyo, Bunkyo-ku, Tokyo 113-0033, Japan
}%


\author{F.-H.~Chang, H.-J.~Lin, D.~J.~Huang, and C.~T.~Chen}
\affiliation{National Synchrotron Radiation Research Center, Hsinchu 30076, Taiwan
}%



\author{G.~Z.~Xing, J.~W.~Seo, C.~Panagopoulos, and T.~Wu}
\affiliation{Division of Physics and Applied Physics, 
School of Physical and Mathematical Sciences, Nanyang Technological University, Singapore 637371
}%

\date{\today}
\begin{abstract}

We have studied the electronic structure and the magnetism of Cu-doped ZnO nanowires, 
which have been reported to show ferromagnetism at room temperature 
[G. Z. Xing ${et}$ ${al}$., Adv. Mater. {\bf 20}, 3521 (2008).], 
by x-ray photoemission spectroscopy (XPS), x-ray absorption spectroscopy (XAS) 
and x-ray magnetic circular dichroism (XMCD). 
From the XPS and XAS results, we find that the Cu atoms are in the ``Cu$^{3+}$" state with 
mixture of Cu$^{2+}$ in the bulk region ($\sim$ 100 nm), 
and that ``Cu$^{3+}$" ions are dominant in the surface region  ($\sim$ 5 nm), 
i.e., the surface electronic structure of the surface region differs from the bulk one. 
From the magnetic field and temperature dependences of the XMCD intensity, 
we conclude that the ferromagnetic interaction in ZnO:Cu NWs comes from the 
Cu$^{2+}$ and ``Cu$^{3+}$" states in the bulk region, 
and that most of the doped Cu ions are magnetically inactive probably 
because they are antiferromagnetically coupled with each other.

\end{abstract}


\maketitle

\section*{I. Introduction}

Diluted magnetic semiconductors (DMSs), where transition-metal (TM) ions are doped into the ZnO hosts, 
are currently receiving intense attention due to the possibility of utilizing both charge and spin degrees of freedom in the same materials, 
allowing us to design new generation spin electronic devices \cite{Furdyna, Ohno, Fukumura}. 
Ferromagnetic ZnO-based DMSs have been predicted to have Curie temperature (T$_{\rm C}$) 
above room temperature (RT) \cite{Dietl, Katayama}, making spintronics applications of ZnO promising. 
However, a recent work \cite{Coey} has suggested that structurally perfect ZnO-based DMSs do not exhibit ferromagnetic order, 
but that not only the magnetic dopants themselves but also vacancies are necessary for ferromagnetism. 

Recently, Cu-doped ZnO (ZnO:Cu) DMS has attracted much attention because of the observations of RT ferromagnetism \cite{Xing, Tian}. 
A recent x-ray magnetic circular dichroism (XMCD) study \cite{Herng} on ZnO:Cu films has suggested a microscopic indirect double-exchange model, 
in which the alignment of localized moments of Cu$^{2+}$ in the vicinity of an oxygen vacancy 
is mediated by the large-sized vacancy orbitals through Cu$^+$-Cu$^{2+}$ ferromagnetic interaction. 
Although this experimental work \cite{Herng} suggests ferromagnetic interaction between the doped
Cu ions via oxygen vacancies and the importance of electron doping for ferromagnetism in ZnO:Cu, 
a recent first-principle calculation \cite{D. Huang} has indicated that $p$-type ZnO:Cu shows
ferromagnetism but $n$-type ZnO:Cu would not have local magnetic moment. Also, an
XMCD study by Thakur ${et}$ ${al}$. \cite{Thakur} has shown that Cu$^{2+}$ and Cu$^{3+}$ ions in ZnO:Cu DMS 
are magnetically active. If the divalent Cu ions are converted from divalent to trivalent due to 
Zn vacancies as pointed by Ganguli ${et}$ ${al}$. \cite{Ganguli}, 
Thakur's work \cite{Thakur} may imply hole-mediated ferromagnetism in ZnO:Cu. 
Also, experimental results  \cite{Xing} using scanning electron microscopy (SEM) and transmission electron microscopy (TEM) 
suggest that structural imperfections due to Cu diffusion yield a greater overall volume occupied by bound magnetic polarons (BMPs), 
promoting the creation and percolation of ferromagnetic regions. 
Thus, no consensus on the origin of the ferromagnetism in ZnO:Cu has been achieved from the electronic structure point of view. 
In order to clarify the issue, fundamental understanding the electronic structure of doped Cu ions in ferromagnetic ZnO:Cu is essential.

In the present work, we have performed x-ray photoemission spectroscopy (XPS), x-ray absorption spectroscpy (XAS) and 
XMCD measurements on ZnO:Cu nanowires (NWs). 
The electronic configuration of the doped Cu ions has been determined by XPS and XAS. 
Note that the line shape and the peak energies of the Cu 2$p$ XPS spectrum are sensitive to the electronic configuration of the Cu ion \cite{Ghijsen}. 
Ferromagnetic interaction between Cu ions has been studied by 
the combination of surface- and bulk-sensitive Cu 2$p$-3$d$ XMCD measurements.

\section*{II. Experimental}

Details of the sample preparation of ZnO:Cu NWs were described in Ref. \cite{Xing}. 
The ZnO NWs were grown by vapor phase transport inside a horizontal quartz tube furnace (Lindberg/Blue Mini-Mite). 
To obtain the sample ZnO:Cu (Cu$\sim$2.2\%) NWs, a 2 nm layer of Cu was sputtered onto the surface of 
the vertically aligned nanowires, followed by a thermal treatment in air at temperatures between 600 and 800 $^\circ$C for 8 h. 
Structural characterization was carried out by x-ray diffraction (XRD), SEM and TEM, 
demonstrating a clear NW phase with no secondary phase \cite{Xing}. 
The ZnO:Cu NW is a column-shaped crystal (not a hollow pillar), and the diameter and length were 120-400 nm and 4 $\mu$m, respectively. 
We refer to the surface layer of  $\sim$5 nm thickness of the NW as the surface region. 
Note that a secondary ion mass spectroscopy study revealed inhomogeneous distribution of Cu atoms in this sample. 

XAS and XMCD measurements were performed at the Dragon Beamline BL-11A of National 
Synchrotron Radiation Research Center (NSRRC), Taiwan. 
The spectra were taken both in the total-electron-yield (TEY: probing depth $\sim$ 5 nm) and the 
total-fluorescene-yield (TFY: probing depth $\sim$ 100 nm) modes. 
The monochromator resolution was $E$/$\Delta$$E$$>$10000 and 
the circular polarization of x-rays was $\sim$ 60\%. 
XPS measurements were performed using the photon energy of $h$$\nu$ =1253.6 eV (Mg-K$\alpha$). 
Photoelectrons were collected using a Scienta SES-100 electron-energy analyzer. 
The energy resolution was about $\sim$ 800 meV.

\section*{III. Results and discussion}

Figure 1 shows the Cu 2$p$ XPS spectra of ZnO:Cu and 
various copper oxides Cu$_2$O \cite{Ghijsen}, CuO \cite{Ghijsen}, and NaCuO$_2$ \cite{Steiner}. 
For ZnO:Cu, one can see a small satellite feature around 945 eV, reflecting the ionic character of doped Cu atoms. 
By comparing the XPS line shape of ZnO:Cu with those of the other oxides, 
it is likely that the electronic structure of the doped Cu ions is similar to that of Cu ions in NaCuO$_2$. 
Thus, we assigned the main peak and the satellite feature of ZnO:Cu 
to 2$\underline{p}$3$d$$^{10}$$\underline{L}$$^2$ ($\underline{p}$ = Cu 2$p$ hole, $\underline{L}$ = ligand hole) and 
2$\underline{p}$3$d$$^9$$\underline{L}$ final states, respectively. 
Although NaCuO$_2$ is known as a formally Cu$^{3+}$ insulator, 
we note that the notation Cu$^{3+}$ does not necessary imply the pure  3$d^8$ configuration 
but includes strong influence of Cu 3$d$$-$O 2$p$ covalency \cite{MizokawaPRL, MizokawaPRB}. 
In Cu$^{3+}$ oxides, in fact, the wave function of the Cu ion fundamental state is a linear combination of 
the 3$d^8$, 3$d$$^{9}$$\underline{L}$ and 3$d$$^{10}$$\underline{L}$$^2$ configurations. 
From XPS results, we suggest that, within the probing depth region of XPS measurement  ($\sim$5-10 nm), 
the doped Cu ions in ZnO:Cu NWs are Cu$^{3+}$ dominated by 3$d$$^{9}$$\underline{L}$ and 3$d$$^{10}$$\underline{L}$$^2$.

\begin{figure}
\includegraphics[width=7cm]{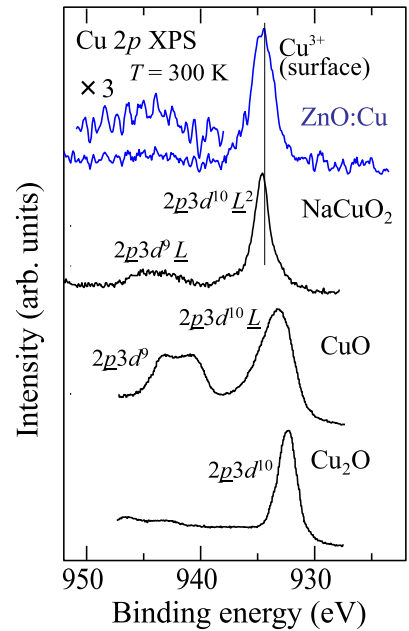}
\caption{(Color online) 
Cu 2$p$ XPS spectra of the ZnO:Cu NWs and other copper oxides 
Cu$_2$O \cite{Ghijsen}, CuO \cite{Ghijsen}, and NaCuO$_2$ \cite{Steiner}.}
\end{figure}

\begin{figure}
\includegraphics[width=7.0cm]{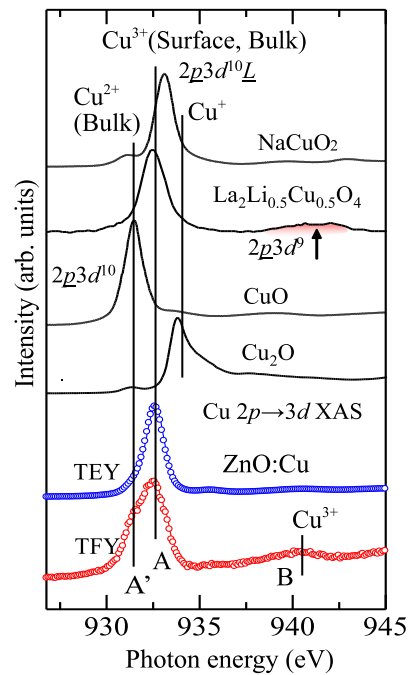}
\caption{(Color online) 
Cu 2$p$$\rightarrow$3$d$ XAS spectra of the ZnO:Cu NWs and other copper oxides 
 Cu$_2$O, CuO \cite{DDSarma}, NaCuO$_2$ \cite{DDSarma} and 
La$_2$Li$_{0.5}$Cu$_{0.5}$O$_4$ \cite{ZHu}.}
\end{figure}

Figure 2 shows the Cu 2$p$$\rightarrow$3$d$ XAS spectra of the ZnO:Cu NWs and 
various copper oxides Cu$_2$O, CuO \cite{DDSarma}, NaCuO$_2$ \cite{DDSarma} and 
La$_2$Li$_{0.5}$Cu$_{0.5}$O$_4$ \cite{ZHu}. 
The main peaks in the spectra of NaCuO$_2$ and La$_2$Li$_{0.5}$Cu$_{0.5}$O$_4$ 
are considered to originate from the 2$\underline{p}$3$d$$^{10}$$\underline{L}$ final state due to the 3$d$$^{9}$$\underline{L}$ ground state. 
The weak feature around 931 eV in the spectrum of NaCuO$_2$ is an extrinsic signal due to the 3$d$$^{9}$ 
ground state of Cu$^{2+}$ impurities \cite{MizokawaPRL, MizokawaPRB}. 
The satellite structure of the XAS spectrum of La$_2$Li$_{0.5}$Cu$_{0.5}$O$_4$ \cite{ZHu} 
is considered to originate from the 2$\underline{p}$3$d$$^{9}$ final state due to the 3$d^8$ ground state. 
Note that the satellite structure cannot be observed in the spectrum of Cu$^{+}$ and Cu$^{2+}$ compounds. 
In the spectrum of NaCuO$_2$ \cite{DDSarma}, 
the satellite intensity is small compared to that of La$_2$Li$_{0.5}$Cu$_{0.5}$O$_4$ \cite{ZHu}, 
reflecting the stronger Cu 3$d$$-$O 2$p$ covalency in NaCuO$_2$ than La$_2$Li$_{0.5}$Cu$_{0.5}$O$_4$. 

In the spectrum of ZnO:Cu, although the Cu 2$p$$_{3/2}$ main peak position of the XAS spectra taken 
both in the surface-sensitive TEY and bulk-sensitive TFY modes is same, the line shape is different. 
This reflects differences in the electronic structure between the bulk and surface regions of the ZnO:Cu NWs. 
Now, we discuss differences in the electronic structure between the surface and bulk regions of the ZnO:Cu NWs. 
By comparing the peak position and line shape of the XAS spectrum of ZnO:Cu taken in the TEY mode with those of the Cu oxides, 
we find that, in the spectrum of ZnO:Cu, 
the main peak A originates from the 2$\underline{p}$3$d$$^{10}$$\underline{L}$ final state due to the 3$d$$^{9}$$\underline{L}$ ground state 
because main peak A is located at the same energy position with the main peak position of  NaCuO$_2$ and La$_2$Li$_{0.5}$Cu$_{0.5}$O$_4$. 
The finding that the doped Cu ions in the surface region ($\sim$ 5 nm) are Cu$^{3+}$, i.e., Cu$^{2+}$ states plus oxygen $p$ holes, 
is consistent with the XPS results. 
On the other hand, in the spectrum taken in the TFY mode, 
the satellite structure B around 940 eV and the feature A' in the leading-edge of the 
Cu 2$p$$_{3/2}$ main peak were observed in addition to the main peak A. 
We find that feature A' is located at the same energy position with the main peak position of  CuO \cite{DDSarma} and 
that the feature originates from the $3d^9$ states, that is, Cu$^{2+}$. 
This indicates the presence of Cu$^{2+}$ states in the bulk region of ZnO:Cu. 
In addition to this, the satellite structure B due to $3d^8$ are observed only in the spectrum of ZnO:Cu taken in the TFY mode. 
This means that Cu 3$d$-O 2$p$ covalency in the surface region is stronger than that in the bulk region. 
From the XAS results, we suggest that the surface electronic structure of the surface region differs from the bulk one and that 
the electronic structures of the doped Cu atoms in the surface and bulk regions of ZnO:Cu are 
predominantly Cu$^{3+}$ and Cu$^{2+}$/Cu$^{3+}$, respectively.

\begin{figure}
\includegraphics[width=9.0cm]{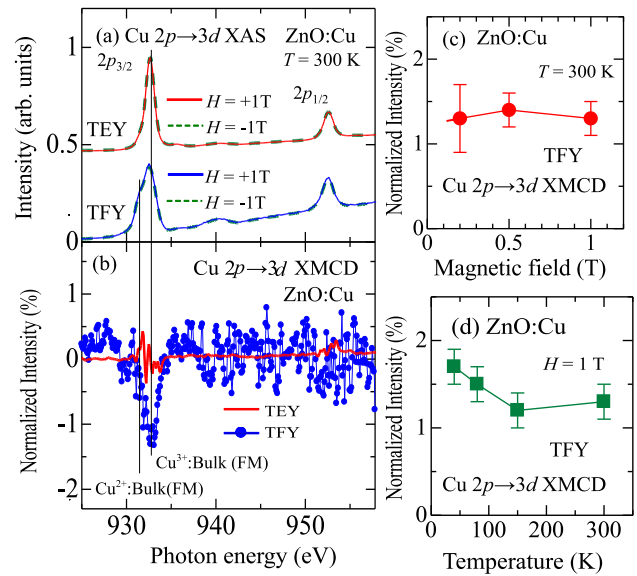}
\caption{(Color online) 
Cu 2$p$$\rightarrow$3$d$ XAS (a) and XMCD (b) spectra of the ZnO:Cu NWs at RT and $H$ =$\pm$1T. 
Magnetic field (c) and temperature (d) dependences of the XMCD intensities.}
\end{figure}

\begin{figure}
\includegraphics[width=9.0cm]{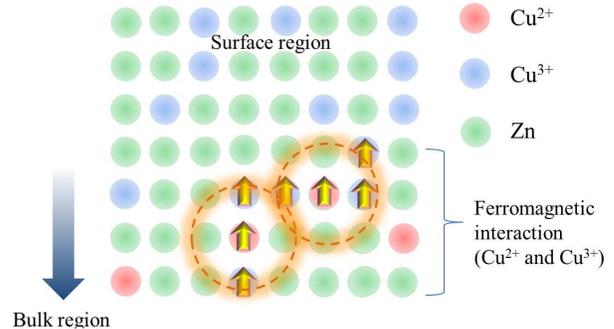}
\caption{(Color online) 
A schematic of magnetic interactions in ZnO:Cu NWs.}
\end{figure}

Finally, we discuss the magnetism of Cu ions in the ZnO:Cu NWs. 
Figures 3(a) and (b) show the Cu 2$p$$\rightarrow$3$d$ XAS and XMCD spectra of the ZnO:Cu NWs, 
respectively, taken both in the TEY and TFY modes at RT and $H$ = $\pm$1 T. 
In Fig. 3(b), the XMCD signal is normalized to the XAS intensity. 
Note that the XMCD strength is free from the diamagnetism of the substrate. 
The XMCD signal taken in the TEY mode is weak or negligible, 
indicating that the Cu$^{3+}$ states in the surface region ($\sim$ 5 nm) are magnetically inactive. 
On the other hand, clear XMCD signal was observed in the spectrum taken in the TFY mode  [see Fig. 3(b)], 
which has the probing depth region of $\sim$ 100 nm and hence probes the entire NW. 
The XMCD signals are located at the same energy positions with the peak positions due to the Cu$^{2+}$ and Cu$^{3+}$ states. 
The Cu 2$p$$\rightarrow$3$d$ XMCD intensity is about $\sim$1.3\%, 
indicating that a small amount of the Cu$^{2+}$/Cu$^{3+}$ states in the bulk region is magnetically active in the ZnO:Cu NWs. 
The fact that the Cu$^{2+}$/Cu$^{3+}$ states are magnetically active is consistent with Thakur's work \cite{Thakur}. 
However, the facts that Cu$^{3+}$ states mainly exist in the surface region ($\sim$ 5 nm) and 
that clear Cu 2$p$$\rightarrow$3$d$ XMCD signals were not observed when taken in the TEY mode disagree with Thakur's work. 
It is likely that the inconsistency can be attributed to the different samples (films ${vs}$ NWs). 
Figures 3(c) and (d) show magnetic field and temperature dependences of the Cu 2$p$$\rightarrow$3$d$ XMCD intensity, respectively, taken in the TFY mode. 
Following a previous XMCD study of DMS \cite{Takeda}, 
we confirmed the presence of ferromagnetism by linear extrapolation of the XMCD intensity to zero field [see Fig. 3(c)]. 
Based on the XMCD intensities taken in the TFY mode, 
we suggest that only a small portion ($\sim$ 1\%) of Cu ions in the bulk region are ferromagnetic (FM). 
On the other hand, we could not observe a clear paramagnetic (PM) component due to the remaining ($\sim$ 99\%) Cu ions, 
because the intensities of the XMCD signals, taken in the TFY mode, were unchanged with magnetic field. 
Therefore, as shown in Fig. 4, we conclude that the FM interaction of ZnO:Cu NWs comes from the Cu$^{2+}$ and Cu$^{3+}$ states in the bulk region and not in the surface region.  
Also, it is likely that, in the bulk region, FM and antiferromagnetic (AFM) interactions coexist and 
that most of the doped Cu ions in the bulk region are antiferromagnetically coupled with each other. 
Our result suggests that the coexistence of the Cu$^{3+}$ states, which are mainly due to the 3$d$$^{9}$$\underline{L}$ ground states, 
i.e., Cu$^{2+}$ states plus oxygen $p$ holes, and the Cu$^{2+}$ states triggers the ferromagnetism in ZnO:Cu NWs.

\section*{IV. Summary}

We have performed XPS, XAS and XMCD measurements on ZnO:Cu NWs. 
From the Cu 2$p$ XPS and Cu 2$p$$\rightarrow$3$d$ XAS results, 
we suggest that the electronic structures of the doped Cu atoms in the surface ($\sim$ 5 nm) and bulk ($\sim$ 100 nm) regions of 
ZnO:Cu are predominantly ``Cu$^{3+}$" and Cu$^{2+}$/``Cu$^{3+}$", respectively. 
XMCD signals due to ferromagnetism were observed at the Cu 2$p$ absorption edge taken in the TFY mode. 
From the magnetic field and temperature dependences of the XMCD intensity, 
we conclude that the FM interaction of ZnO:Cu NWs comes from the Cu$^{2+}$ and ``Cu$^{3+}$" states in the bulk region 
and that most of the doped Cu ions are magnetically inactive probably 
because they are antiferromagnetically coupled with each other. 
Our result suggests that the hole doping in ZnO:Cu NWs plays an important role in the ferromagnetism.

\section*{Acknowledgement}
The work was supported by a Grant-in-Aid for Scientific Research
(S22224005) from JSPS, and The National Research Foundation of Singapore.


\begin{thebibliography}{10}

\bibitem{Furdyna}
J. K. Furdyna, 
J. Appl. Phys. {\bf 64},  R29  (1988).

\bibitem{Ohno}
H. Ohno, Science {\bf 281},  951  (1998).

\bibitem{Fukumura}
H. Toyosaki, T. Fukumura, Y. Yamada, K. Nakajima, T. Chikyow, T. Hasegawa, H. Koinuma, and M. Kawasaki, 
Nature Mater. {\bf 3},  221  (2004).

\bibitem{Dietl}
T. Dietl, H. Ohno, F. Matsukura, J. Cibert, and D. Ferrand, 
Science {\bf 287},  1019  (2000).

\bibitem{Katayama}
K. Sato and H. Katayama-Yoshida, 
Jpn. J. Appl. Phys. {\bf 40},  L334  (2001).

\bibitem{Coey}
J. M. D. Coey and S. A. Chambers, 
MRS Bulletin  {\bf 33}, 1053 (2008).

\bibitem{Xing}
G. Z. Xing, J. B. Yi, J. G. Tao, T. Liu, L. M. Wong, Z. Zhang, G. P. Li, S. J. Wang, 
J. Ding, T. C. Sum, C. H. A. Huan, T. Wu, Adv. Mater.  {\bf 20}, 3521 (2008). 

\bibitem{Tian}
Y. F. Tian, Y. F. Li, M. He, I. A. Putra, H. Y. Peng, B. Yao, S. A. Cheong, and T. Wu, 
Appl. Phys. Lett. {\bf 98}, 162503 (2011).

\bibitem{Herng}
T. S. Herng, D.-C. Qi, T. Berlijn, J. B. Yi, K. S. Yang, Y. Dai, Y. P. Feng, 
I. Santoso, C. Sanchez-Hanke, X. Y. Gao, Andrew T. S. Wee, W. Ku, 
J. Ding, A. Rusydi, 
Phys. Rev. Lett. {\bf 105}, 207201 (2010).

\bibitem{D. Huang} 
D. Huang, Y.-J. Zhao, D.-H. Chen, and Y.-Z. Shao,
Appl. Phys. Lett. {\bf 92}, 182509 (2008).

\bibitem{Thakur}
P. Thakur, V.  Bisogni, J. C.  Cezar, N. B.  Brookes, G. Ghiringhelli, S. Gautam, 
K. H. Chae, M. Subramanian, R. Jayavel, K. Asokan, 
J. Appl. Phys. {\bf 107}, 103915, (2010).

\bibitem{Ganguli}
N. Ganguli, I. Dasgupta and B. Sanyal, 
Appl. Phys. Lett. {\bf 94}, 192503 (2009).

\bibitem{Ghijsen}
J. Ghijsen, L. H. Tjeng, J. van Elp, H. Eskes, J. Westerink, 
G. A. Sawatzky, and M. T. Czyzyk, 
Phys. Rev. B {\bf 38}, 11322 (1988).

\bibitem{Steiner}
P. Steiner, V. Kmsinger, I. Sander, B. Siepart, S. H fner, C. Politic, 
R. Hoppe and H. P. Muller, 
Z. Phys. B {\bf 67}, 497 (1987).

\bibitem{MizokawaPRL} 
T. Mizokawa, H. Namatame, A. Fujimori, K. Akeyama, 
H. Kondoh, H. Kuroda, and N. Kosugi, 
Phys. Rev. Lett. {\bf 67}, 1638 (1991).

\bibitem{MizokawaPRB} 
T. Mizokawa, A. Fujimori, H. Namatame, 
K. Akeyama, and N. Kosugi, 
Phys. Rev. B.  {\bf 49}, 7193, (1994).


\bibitem{DDSarma}
D. D. Sarma, O. Strebel, C. T. Simmons, U. Neukirch, and G. Kaindl, R. Hoppe, and H. P. M\"uller, 
Phys. Rev. B  {\bf 37}, 9784 (1988).

\bibitem{ZHu}
Z. Hu, G. Kaindl, S.A. Warda, D. Reinen, F. M. F. de Groot, B. G. M\"{u}ller, 
Chem. Phys. {\bf 232}, 63 (1998).


\bibitem{Takeda} 
Y. Takeda, M. Kobayashi, T. Okane, T. Ohkochi, J. Okamoto, Y. Saitoh, 
K. Kobayashi, H. Yamagami, A. Fujimori, A. Tanaka, J. Okabayashi, 
M. Oshima, S. Ohya, P. N. Hai, and M. Tanaka, 
Phys. Rev. Lett. {\bf 100}, 247202 (2008).



\end{thebibliography}
\end{document}